\documentclass{aa}  

\usepackage{graphicx}
\usepackage{lipsum}  
\usepackage{mathtools}
\usepackage{siunitx}
\usepackage[bottom]{footmisc}
\usepackage{tablefootnote}
\usepackage[flushleft]{threeparttable}
\usepackage{adjustbox}
\usepackage{subcaption}
\usepackage{float}

\usepackage{txfonts}
\begin{document} 

   \title{Is there a dwarf galaxy satellite-of-satellite problem in $\Lambda$CDM?}
   \author{Oliver Müller
          \inst{1}
        \and 
          Nick Heesters\inst{1}
          \and
          Helmut Jerjen\inst{2}
                    \and
          Gagandeep Anand\inst{3}
          \and 
          Yves Revaz\inst{1}
          }
\titlerunning{A satellite-of-satellite problem?}
   \institute{Institute of Physics, Laboratory of Astrophysics, Ecole Polytechnique F\'ed\'erale de Lausanne (EPFL), 1290 Sauverny, Switzerland\\
              \email{oliver.muller@epfl.ch}
         \and
             Research School of Astronomy and Astrophysics, Australian National University, Canberra, ACT 2611, Australia\\
             \email{helmut.jerjen@anu.edu.au}
        \and
   Space Telescope Science Institute, 3700 San Martin Drive, Baltimore, MD 21218, USA
             }
   \date{Received September tbd; accepted tbd}

 
  \abstract
   {
    Dark matter clusters on all scales, therefore it is expected that even substructure should host its own substructure. Using the Extragalactic Distance Database, we searched for dwarf galaxy satellites of dwarf galaxies, i.e. satellite-of-satellite galaxies, corresponding to these substructures-of-substructure. Going through Hubble Space Telescope data of 117 dwarf galaxies, we report the discovery of a previously unknown dwarf galaxy around the ultra-diffuse M96 companion M96-DF6 at 10.2\,Mpc in the Leo-I group. We confirm its dwarf galaxy nature as a stellar overdensity. Modelling its structural parameters with a growth curve analysis, we find that it is an ultra-faint dwarf galaxy with a luminosity of 1.5 $\times$ 10$^5$\,L$_\odot$, which is 135 times fainter than its host. 
    Based on its close projection to M96-DF6 it is unlikely that their association occurs just by chance. We compare the luminosity ratio of this and three other known satellite-of-satellite systems with results from two different cosmological sets of $\Lambda$CDM simulations. For the observed stellar mass range of the central dwarf galaxies, the simulated dwarfs have a higher luminosity ratio between the central dwarf and its first satellite ($\approx$10'000) than observed ($\approx$100), {excluding the LMC/SMC system}. No simulated dwarf analog at these observed stellar masses has the observed luminosity ratio. This cannot be due to missing resolution, because it is the brightest subhalos that are missing. This may indicate that there is a satellite-of-satellite {(SoS)} problem for $\Lambda$CDM {in the stellar mass range between 10$^6$ and 10$^8$\,M$_\odot$ -- the regime of the classical dwarf galaxies}. {However, simulated dwarf models at both a lower ($<$10$^6$\,M$_\odot$) and higher ($>$10$^8$\,M$_\odot$) stellar mass have comparable luminosity ratios. For the higher stellar mass systems, the LMC/SMC system is reproduced by simulations, for the lower stellar masses, no observed satellite-of-satellite system has been observed to date. More observations and simulations of satellite-of-satellite systems are needed to assess whether the luminosity ratio is at odds with $\Lambda$CDM.}
   }

   \keywords{Galaxies: dwarf, Galaxies: groups: individual: M\,96/Leo-I group, Galaxies: luminosity function, mass function
               }

   \maketitle
%

\section{Introduction}
In a Lambda Cold Dark Matter ($\Lambda$CDM) cosmology, matter clusters on all scales -- from the biggest clusters of galaxies down to the smallest stars. Every dark matter halo possesses dark matter substructure. For galactic sized dark matter halos, the well-known substructure is represented by dwarf galaxies swarming their giant host galaxies. The connection between substructure galactic halos has been extensively used to test $\Lambda$CDM predictions \citep[e.g., ][]{1999ApJ...524L..19M,2005A&A...431..517K,2010MNRAS.401.1889L,2011MNRAS.415L..40B,2016MNRAS.457.1931S,2017ApJ...850..132P,2019ApJ...870...50J,2020A&A...644A..91M}. 

While the study of the substructure is highly dependent on the resolution of the dark matter and baryonic particles, modern simulations are capable of simulating the faintest known dwarf galaxies (e.g. \citealt{2018A&A...616A..96R,2019MNRAS.490.4447W}) with some caveats, such as the size of the simulation box. This makes it possible to predict and study the substructures-of-substructure. 

The closest satellite within the Milky Way halo to study its own satellite population is the LMC. The LMC has one bright satellite -- the SMC -- as well as potentially several ultra-faint dwarf companions \citep{2020MNRAS.495.2554E,2020ApJ...893..121P,2022A&A...657A..54B,2022ApJ...940..136P}. 
\citet{2017MNRAS.472.1060D} calculated  the expected number of bright ($M_*>$10$^4$\,M$_\odot$) and faint ($M_*<$10$^4$\,M$_\odot$) satellites  associated with the LMC using abundance matching and simulations. They noted that observations may be in tension with the LMC system, because four to eight such bright galaxies are predicted, but only one is known. This indicates that there {may be} something off in the abundance of satellite systems. {However, it has also been argued that the classical dwarf Carina may be associated with the LMC  based on the orbital angular momentum, which would lessen this tension \citep{2019MNRAS.489.5348J,2020ApJ...893..121P,2022A&A...657A..54B}.}

To extend the {studies of LMC satellites}, \citet{2017MNRAS.472.1060D} made predictions for the number of satellites more massive than 10$^5$\,M$_\odot$ for several nearby LMC-like galaxies. \citet{2020A&A...644A..91M} searched for dwarf galaxies around two of those galaxies, as well as added data for four additional galaxies from the literature \citep{2009ApJ...705..758M,2016ApJ...828L...5C,2020ApJ...891..144C}. They found that the observed satellite numbers tend to be on the lower side of the expected range. Either there is an over-prediction of luminous subhalos in $\Lambda$CDM or observational biases may underestimate the true number of satellites, or a combination of both. The former is reminiscent to the too-big-to-fail problem \citep{2011MNRAS.415L..40B}, the latter could arise due to e.g. projections of bright foreground stars or background galaxies at the position of an -- overlooked -- dwarf galaxy. 

To further study the abundance of substructure  it is therefore imperative to survey different environments and test whether there is a problem in $\Lambda$CDM or it is due to observational uncertainties. In this article, we present the discovery of such a satellite system in the nearby Leo-I group. We further discuss the luminosity ratios of other observed satellites and their satellites and what is expected in $\Lambda$CDM. 


\section{Data and Methods}
\label{data}

In this section, we first present the data and search for dwarf galaxies on which this paper is based on and then perform surface and point source photometry and estimate the stellar over-density of the newly discovered dwarf galaxy.

\begin{figure*}[ht]
    \centering
    \includegraphics[width=\linewidth]{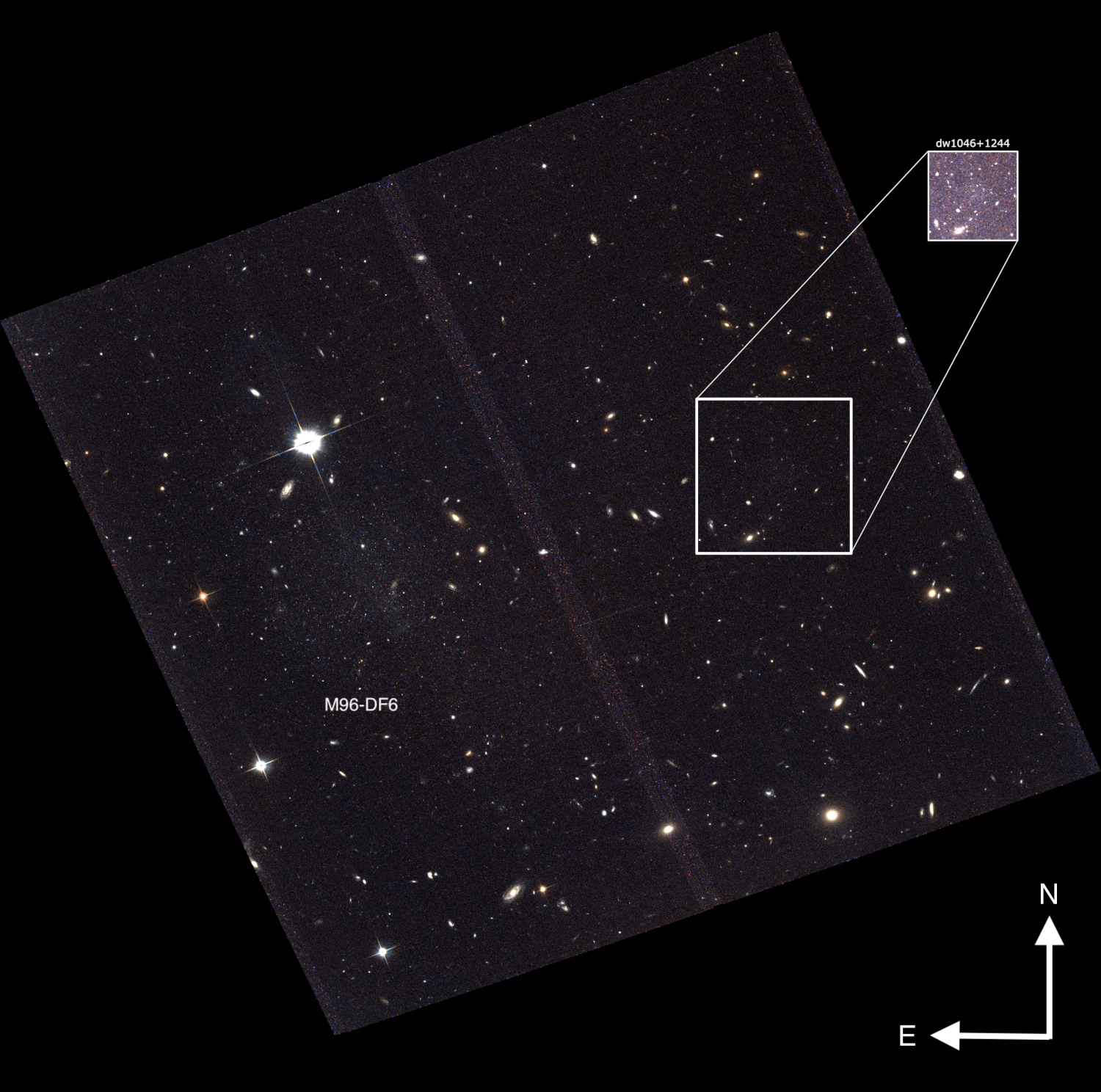}

    \caption{{Pseudo-color image created from HST/ACS F606W and F814W images showing the {ultra diffuse} galaxy M96-DF6 (below the bright foreground star to the left). The large square box shows the location of the new satellite {dw1046+1244} {and is 23\,arcsec per side}. To visualize this faint dwarf we blurred the box with a Gaussian filter and adjusted the brightness curve (small white box). North is up and East is to the left. Image from the HST Program 14644.}}
    \label{fig:pv}
\end{figure*}

\subsection{Search for dwarf galaxy satellites}

To search for satellites of dwarf galaxies, we used the Extragalactic Distance Database's\footnote{https://edd.ifa.hawaii.edu/, last accessed: 10.11.2022.} (EDD, \citealt{2009AJ....138..323T}) CMDs/TRGB catalog \citep{2021AJ....162...80A} . We downloaded HST images of 117 dwarf galaxies having a distance estimate larger than 8\,Mpc and carefully inspected them by eye. We choose 8\,Mpc as a minimal distance so that a potential satellite will fall on the field of view of ACS. With 202 $\times$ 202 arcsec$^2$ ACS covers at least a radial distance of 4\,kpc  
 although we note that the targeted dwarf galaxy often was centered on one of the two CCDs. Hence the area around the dwarf was not covered uniformly to the same distance in all directions.

For PGC\,4689210, a dwarf galaxy in the M96 group -- also known as Leo I 09 \citep{2002MNRAS.335..712T}, or NGC3384-DF6, M96-DF6 \citep{2018ApJ...868...96C} -- we detected a faint {stellar over-density, hereafter called dw1046+1244, close to the galaxy (see Fig.\ref{fig:pv}). It has a similar color and surface brightness granulation, indicating that it may be at the same distance as M96-DF6. In the HST image, M96-DF6 is barely resolved into stars.  M96-DF6 is an ultra-diffuse galaxy in the Leo-I group with a tip of the red giant branch (TRGB) distance of 10.2$\pm$0.3 Mpc \citep{2018ApJ...868...96C}.

\subsection{Point source photometry}

We perform point-spread function (PSF) photometry with the DOLPHOT software package \citep{Dolphin00, DOLPHOT}. Using the parameters used in the Extragalactic Distance Database’s pipeline \citep{Anand21b} which are based on the DOLPHOT manual’s recommended parameters\footnote{http://americano.dolphinsim.com/dolphot/dolphotACS.pdf}, we perform the photometry on the \textit{*.flc} images which are corrected for losses due to imperfect charge transfer efficiency. We use the F814W drizzled (\textit{*.drc}) image as a reference frame for both alignment and source detection.  Using the photometry measured from the individual frames, DOLPHOT then provides combined photometry and uncertainties for each source overall. The output photometry is supplied in the Vega magnitude system.

\begin{figure}[ht]
    \centering
    \includegraphics[width=0.49\linewidth]{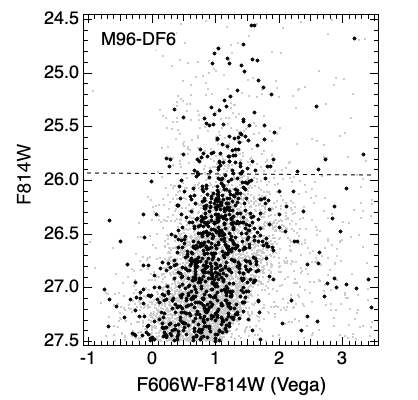}
     \includegraphics[width=0.45\linewidth]{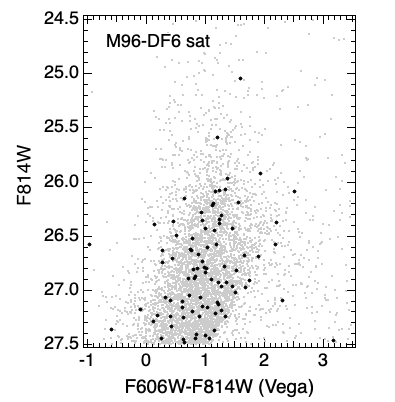}
    \caption{{CMDs for M96-DF6 and the M96-DF6 satellite dw1046+1244. 
    The grey dots are all point sources passing the photometric quality cuts detected over the entire FoV. The dashed line in the left graph indicates the magnitude of the TRGB derived by \citet{2018ApJ...868...96C}.}}
    \label{fig:cmds}
\end{figure}

{Only objects that matched in both F606W, F814W pass-bands and with global quality flag=1, crowding parameter $<0.5$, chi-square parameter $<1.5$, $mag err<0.4$ and sharpness parameter between $-1.5$ and 1.5 were retained. Restricting the sharpness and chi parameters excludes background galaxies as well as any remaining blemishes. The resulting color-magnitude diagrams are shown in Fig. \ref{fig:cmds}. }

\subsection{Surface photometry}

\begin{figure}[ht]
    \centering
    \includegraphics[width=\linewidth]{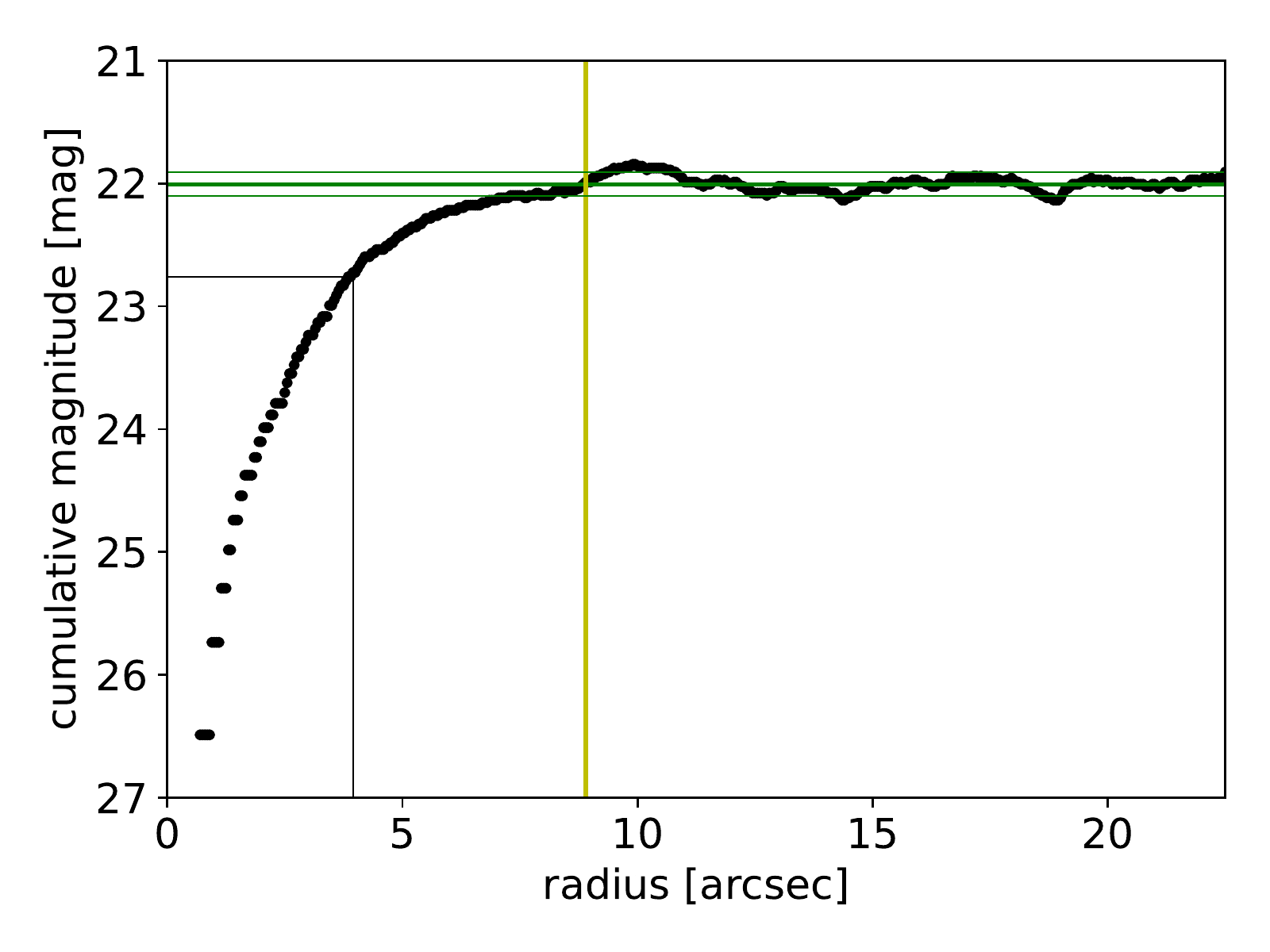}
    \caption{The magnitude growth curve in the F606W band of the newly discovered dwarf galaxy dw1046+1244. The horizontal thick green line indicates the estimated total magnitude, the two thin green lines correspond to $\pm$2 standard deviations around it. The vertical yellow line indicates the radius where the magnitude reaches the background. The two black lines correspond to the distance where the curve reaches half the flux, i.e. the estimation of the effective radius.}
    \label{fig:curve}
\end{figure}

The newly detected dwarf galaxy is faint and has an extremely diffuse stellar distribution in the HST images, which makes surface photometry measurements difficult. Modelling the galaxy light with GALFIT \citep{2002AJ....124..266P} out of the box did not produce meaningful results. Therefore, we first used a growth curve analysis to measure the photometric parameters of the dwarf. This two-step approach has been successfully applied on dwarf galaxies in nearby groups \citep[e.g., ][]{2015A&A...583A..79M,Muller2017}.
For that to work, we mask all sources using the python implementation of Source Extractor \citep{1996A&AS..117..393B} -- SEP \citep{SEP} -- with a threshold of 5$\sigma$. This threshold was chosen via trial and error. Source Extractor generates a source map which we can use as a mask. Because, however, Source Extractor is struggling to detect the low-surface brightness edges of objects, we dilate this mask with a circular kernel of radius 10 pixels. Then we calculate the radial cumulative curve. Masked patches of the sky were replaced with the median estimate of the global background on the masked image. To estimate the local background of the growth curve, the curve is varied until the outer part -- which corresponds to the background -- becomes flat. We find the best value for the local background by stepping through different values and fit a line for pixel values which are well outside of the visible galaxy profile (12.5 to 22.5 arcsec). The local background value which results in a slope of zero gives the best estimate. Because the growth curve is a cumulative sum, the total collected light from the galaxy corresponds to where the curve becomes flat. We find this sum by estimating the median value of the background region (i.e. within 12.5 to 22.5 arcsec). The error is given by two times the standard deviation withing this region {(0.11\,mag)}. We then estimate at which radius the galaxy reaches half of its cumulative sum. This corresponds to the effective radius. Its uncertainty is derived from the uncertainty of the cumulative sum {(0.5\,arcsec)}. See Fig.\,\ref{fig:curve} for a plot of the growth curve. We note that this method of structural parameter estimation is independent of models such as a S\'ersic profile. {The effective surface brightness is calculated from the apparent magnitude and effective radius.}

To test our implementation of the growth curve fitting and estimate the overall uncertainty, we injected 200 artificial dwarf galaxies into the HST data. The injected dwarfs were made to resemble the detected object, with their apparent magnitudes and effective radius uniformly distributed within the 2$\sigma$ measurement uncertainty and between 3 and 5\,arcsec for the effective radius, respectively. For each such dwarf, we repeated the growth curve fitting and estimated the difference between the injected and the extracted structural parameters, namely the apparent magnitude and the effective radius. For the apparent magnitude, we find an offset of $-$0.09\,mag and a standard deviation of 0.11\,mag. For the effective radius, we find  an offset of 0.3\,arcsec and a standard deviation of 0.4\,arcsec.  This is likely coming from underestimating the flux due to the masking and can be corrected by making the dwarf galaxy brighter and larger by these estimated values. {We correct the apparent magnitude and effective radius by these values, and derive the final photometric errors as a root sum square of the uncertainty of the growth curve and the offset from the artificial galaxy experiment. The uncertainty in the effective surface brightness is calculated from a Gaussian error propagation of the apparent magnitude and effective radius.
The structural parameters and the final errors are listed in Table\,\ref{tab:dwarf}.}

\begin{figure*}[ht]
    \centering
    \includegraphics[width=\linewidth]{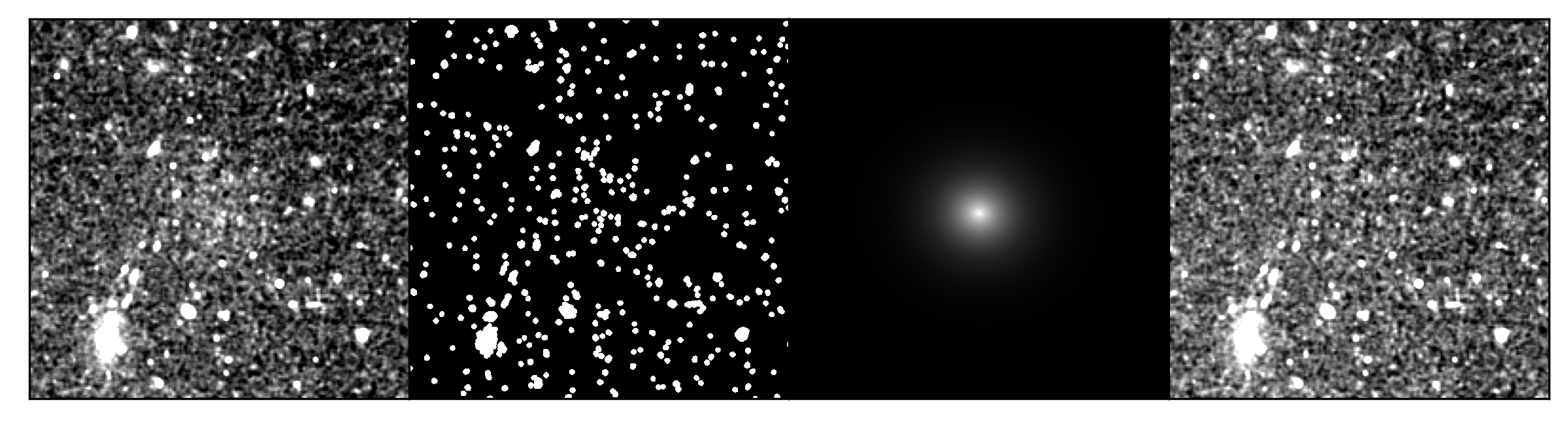}
    \caption{GALFIT modelling of the newly discovered dwarf galaxy dw1046+1244 in the F606W band. The images were smoothed with a Gaussian kernel and binned. From left to right: The snapshot of the dwarf, mask of all sources with a threshold of 5$\sigma$ identified by Source Extractor, best GALFIT model, and residual image.}
    \label{fig:galfit}
\end{figure*}

As a last step, we use GALFIT to model the galaxy. We fix the effective radius, as well as the ellipticity ($0.9<e<1.0$) and an exponential profile (S\'ersic $n=1$). The results of the galfit modelling are presented in Fig.\,\ref{fig:galfit}. The dwarf galaxy light is well subtracted.

\begin{table}[ht]
\caption{The structural properties of the newly discovered dwarf galaxy dw1046+1244.}             
\renewcommand{\arraystretch}{1.3}
\centering                          
\begin{tabular}{l c}        
\hline\hline                 
 & dw1046+1244  \\    
\hline      \\[-2mm]                  
RA (J2000) &  10:46:45.96 \\
Dec (J2000) &  $+$12:44:59.5 \\
$m_{F606W}$ (mag) & 21.92 $\pm$ 0.14  \\
$r_{eff}$ (arcsec)&  4.2 $\pm$ 0.6 \\
$\mu_{eff, F606W}$ (mag arcsec$^{-2}$)& 27.0  $\pm$ 0.3\\
\\
$M_{F606W}$ (mag) & $-$8.1 $\pm$ 0.3  \\
$r_{h}$ (pc)& 210$\pm$ 29  \\
$L_*$ (10$^5$ $L_\odot$)& 1.5$\pm$ 0.3   \\
\\
$M_*$ (10$^5$ $M_\odot$)&  2.3$\pm$ 0.3 \\
\hline
\end{tabular}
\tablefoot{The parameters $m_{F606W}$ and $r_{eff}$ have been corrected {by -0.09\,mag} for systematics calculated in our artificial galaxy tests (see text).  To estimate $M_{F606W}$, $r_{h}$, and $L_*$ we adopted a distance of 10.2\,Mpc, as measured for M96-DF6, and $M_*$ is based on a mass-to-light ratio of 2, typical for dwarf galaxies (e.g., \citealt{Muller2021a}).}
\label{tab:dwarf}
\end{table}

\subsection{Stellar over-density}

\begin{figure*}[ht]
    \centering
    \includegraphics[width=\linewidth]{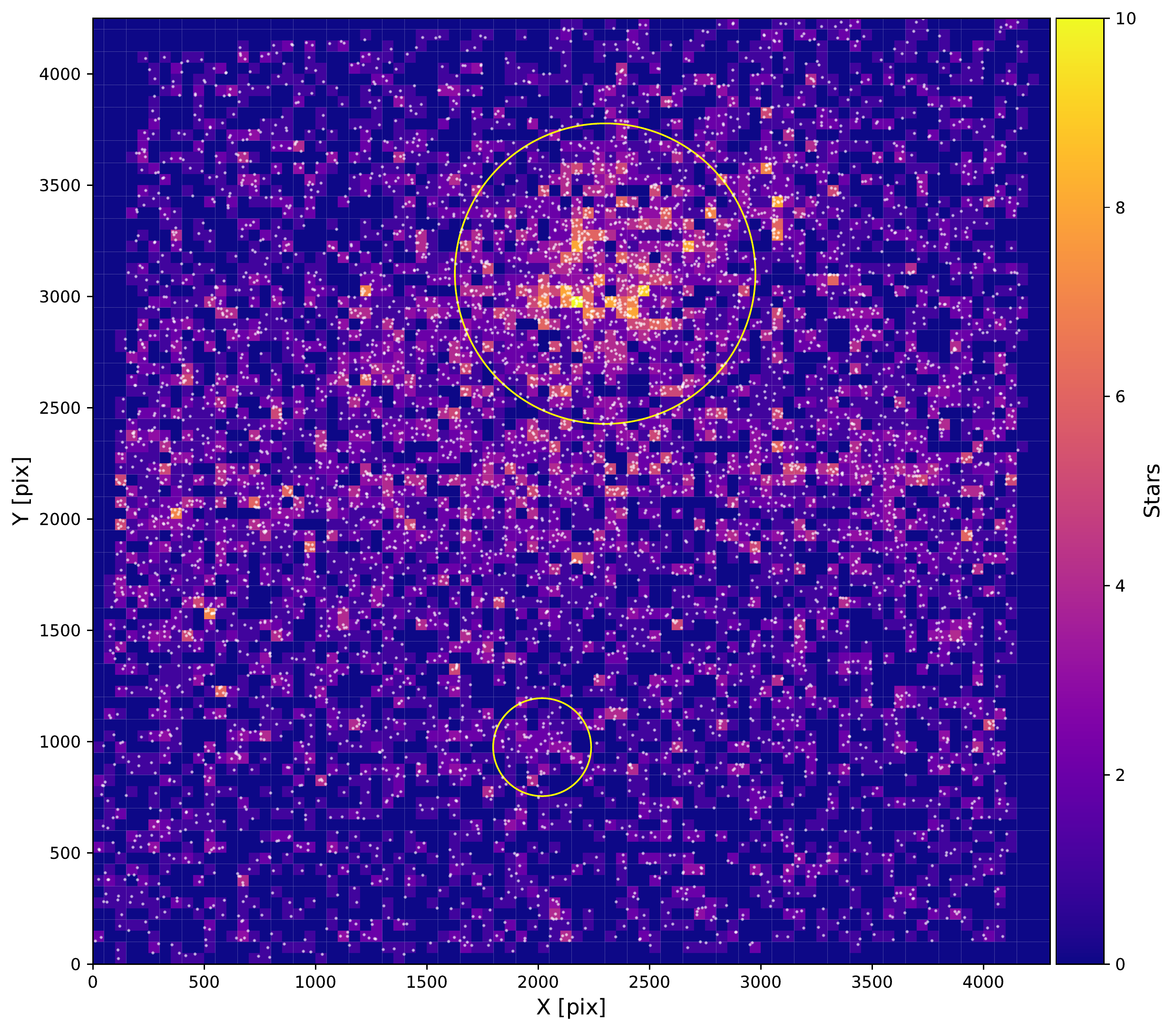}
    \caption{The density map of presumed RGB stars (yellow dots), using a binning of with $50\,x\,50$\,pix$^2$. The yellow circles indicate where the stellar density of the dwarf galaxies drop to the background. The colors of the bins indicate the stellar density -- dark blue meaning no stars and yellow meaning 10 stars.}
    \label{fig:2Dhist}
\end{figure*}

To quantify the stellar over-density associated with the newly discovered ultra-faint dwarf galaxy, we compare the local stellar density in the dwarfs location with the mean stellar density in its environment. To do this we split the field and only consider point sources passing the photometry quality cuts which reside on the chip of the dwarf. Starting at the center of the dwarf we draw a small circle around it and compute the stellar density in this region. Then we normalize it with the mean stellar density on this side of the field (i.e. on the chip where the dwarf galaxy resides). We increase the radius in steps of 20\,pix and repeat the calculation for the annuli with radii $r_{inner} = r_{i}$ and $r_{outer} = r_{i+1}$. 
We note a stellar over-density of $3$ -- corresponding to a 3.4$\,\sigma$ signal -- at the location of the dwarf when compared to its environment. The density drops down to the background level at a radius of $r$ = 220\,pix which corresponds to 10.8\,arcsec. This is similar to the break we find in our growth curve analysis (see Fig.\,\ref{fig:curve}). 
We repeat this test for the dwarf M96-DF6 
in its environment and obtain a stellar over-density radius of  33.1\,arcsec. 
In Figure \ref{fig:2Dhist} we show a 2D histogram of the field with $50\,x\,50$\,pix$^2$ bins and indicate the radii of the stellar over-densities of the two dwarf galaxies.

\section{Discussion}

\subsection{Scaling relation}

\begin{figure}[ht]
    \centering
    \includegraphics[width=\linewidth]{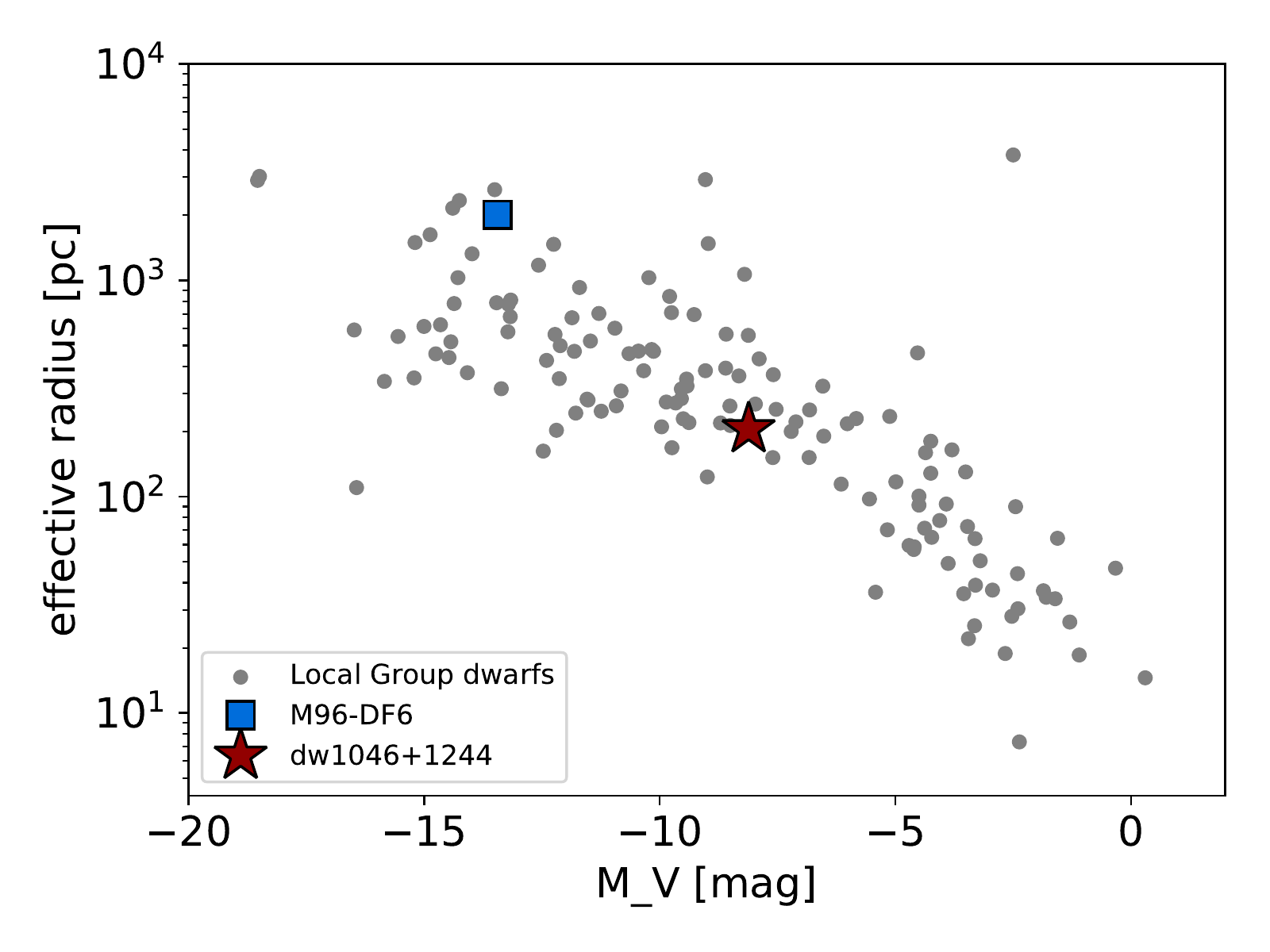}
    \caption{The luminosity-effective radius relation, defined by the Local Group dwarf galaxies (gray dots). The blue square is the ultra-diffuse galaxy M\,96-DF6, the red star corresponds to the newly discovered ultra-faint dwarf dw1046+1244.}
    \label{fig:scaling}
\end{figure}

Dwarf galaxies follow specific scaling relations, such as the luminosity-effective radius relation \citep{2008ApJ...684.1075M,2021MNRAS.506.5494P} and can therefore be used to assess the nature of the detected object (see e.g. Fig\,11 in \citealt{2017A&A...597A...7M}). In Fig.\,\ref{fig:scaling} we present the luminosity-effective radius relation defined by the Local Group dwarf galaxies, as compiled by \citet{2012AJ....144....4M}. As expected, both the dwarf galaxy M96-DF6 and its newly discovered satellite follow this relation. The dwarf galaxy dw1046+1244 is similar to AndXXX \citep{2016ApJ...833..167M}, also known as CassII, with $M_V=-8.0$\,mag and $r_h=260$\,pc. This dwarf has a velocity dispersion of 11.8$^{+7.7}_{-4.7}$ and is  dark matter dominated \citep{2013ApJ...768..172C}. If dw1046+1244 has a similar dark matter halo, such a velocity dispersion can be measured with modern facilities such as Keck (see e.g. \citealt{2019ApJ...874L..12D}) {or the VLT (e.g. \citealt{2019A&A...625A..76E})}. 

With a solar mass of 2 $\times$ 10$^5$ $M_\odot$, the dwarf galaxy is at the threshold of the ultra-faint dwarf galaxy regime ($<$ 10$^5$ $M_\odot$ according to \citealt{2017ARA&A..55..343B}). While this is not a physical distinction to dwarf spheroidals, it denotes the class of objects in the Local Group discovered in the CCD era (e.g. \citealt{2005ApJ...626L..85W,2007ApJ...654..897B,2010ApJ...712L.103B,2009ApJ...705..758M,2015ApJ...804L..44K,2015ApJ...808L..39K}), and separates them from the classical dwarfs, such as Fornax or Sculptor \citep{1938Natur.142..715S}, discovered on photographic plates. Because dw1046+1244 is extremely faint and shallow in surface brightness, we find it appropriate to call it an ultra-faint dwarf galaxy.

\subsection{Satellite of a satellite}

The Leo-I galaxy group has been surveyed by many research teams. \citet{2018A&A...615A.105M} presented the results from a wide search for dwarf galaxies using SDSS. Extrapolating their luminosity function from their Fig.\,6, we expect roughly 160 dwarf galaxies in the environment of the Leo-I group with a luminosity of -8\,mag or more. So what is the likelihood to find two dwarf galaxies within the ACS field of view just by chance? Here we have to be careful, because ACS was pointing towards one dwarf. So the question is rather, what is the probability of finding one additional dwarf galaxy in an ACS image by chance. Assuming the dwarf galaxies are uniformly distributed {within a 4 degree radius} of M\,96, they are distributed within 50.6 deg$^2$. The ACS instrument has a field-of-view of 202 $\times$ 202 arcsec$^2$. The M\,96 footprint is therefore 16'000 times larger than the ACS field-of-view. With 160 potential dwarf galaxies, the probability that one dwarf galaxies falls into the area of the ACS footprint is one percent. It is therefore unlikely, but not impossible, that we see the two of them in the same image just by chance. For now, we consider dw1046+1244 to be a satellite of M\,96-DF6. 
As a caveat: the previous estimate is simplified. Rather than using a uniform distribution we would need to assume a three dimensional radial profile and calculate the cone defined by ACS's field-of-view. However,  ultimately, the satellite nature needs to be confirmed with velocity measurements, so any contamination estimation is only a rough guideline.

The ultra-diffuse galaxy M96-DF6  has a stellar luminosity of 2$\times$10$^7$\,L$_\odot$. This is $\approx$100 more luminous than its presumed satellite galaxy. It is separated from dw1046+1244 by only 1.75\,arcmin, which corresponds to 5.4\,kpc at their measured distance.  \citet{2014ApJ...795L..35C} reported the first discovery of a  satellite of a satellite system in a group of galaxies outside of the Local Group. The two galaxies, Cen\,A-MM-dw1 and Cen\,A-MM-dw2, reside in the nearby Centaurus Group and have stellar luminosities of 2.0$\times$10$^6$\,L$_\odot$ and 2.0$\times$10$^5$\,L$_\odot$, respectively, i.e. a luminosity ratio of 10. The host,  Cen\,A-MM-dw1, has an effective radius of 1.4\,kpc and borders therefore to the ultra-diffuse galaxy regime. The satellite,  Cen\,A-MM-dw2, is an ultra-faint dwarf galaxy similar to the dwarf galaxy reported here, but is slightly larger (360$\pm$80\,pc). They are separated by 3\,kpc. This system is therefore highly similar to the one found here. It is especially intriguing that both M96-DF6 and Cen\,A-MM-dw1 can be considered ultra-diffuse galaxies, which have received attention due to measurements showing that some of them either host a very massive dark matter halo \citep{2016ApJ...828L...6V,2021ApJ...923....9M} or a small -- if any -- dark matter halo \citep{2018Natur.555..629V,2019A&A...625A..76E,2019ApJ...874L..12D}.

Another satellite-of-satellite system was reported by \citet{2018MNRAS.474.3221M}. This system --  LV J1157+5638 and LV J1157+5638 sat -- was also discovered being on the same ACS field. They have luminosities of  1.7$\times$10$^7$\,L$_\odot$ and  4.8$\times$10$^5$\,L$_\odot$, respectively, which is a factor of 36. They are separated by 3.9\,kpc from each other. Again, this is similar to the properties found here.

In the Centaurus group, two other potential galaxy pairs were reported by \citet{2017A&A...597A...7M}, -- dw1243-42/dw1243-42b and dw1251-40/dw1252-40, with projected separations of 1.6\,kpc and 2.9\,kpc, respectively. While the first pair has similar luminosities to each other (a magnitude difference of only 0.5\,mag), the second pair has a luminosity ratio of 16 with  4.5$\times$10$^5$\,L$_\odot$  and 7.3$\times$10$^6$\,L$_\odot$. However, these two pairs both lack accurate distance estimates, so it is not clear if they reside in the same group. We compile all these systems in Table \ref{tab:observed}.

\begin{table}[ht]
\caption{The satellite-of-satellite systems observed in the nearby universe.}             
\centering                          
\begin{tabular}{l l l l l}        
\hline\hline                 
 Dwarf \& 1. satellite  & L$_{*}$  &  L$_{*, 1. satellite}$ & sep. & ratio\\    
 &  10$^6$ L$_{\odot}$ & 10$^5$ L$_{\odot}$ & kpc & \\    
\hline      \\[-2mm]                  
LMC \& &         1486 &    449 &   20 & 3 \\
SMC\\
M96-DF6 \&  &         20.4 &    1.5 &   5 &  135 \\
dw1046+1244  \\
LV J1157+5638 &         17.2 &    0.5  &   4 & 35 \\
LV J1157+5638 sat\\
Cen\,A-MM-dw1 &         2.0 &    0.2 &   3 & 10 \\
Cen\,A-MM-dw2 \\
\\
dw1252-40 \& &         7.3 &    4.5 &   3 & 16 \\
dw1251-40\\
dw1243-42b \& &         2.7 &    16.6 &   2 & 2 \\
dw1243-42 \\
\hline
\end{tabular}
\tablefoot{The references for the discovery of the satellite pairs are: M96-DF6/dw1046+1244  (this work), LV J1157+5638/LV J1157+5638 \citep{2018MNRAS.474.3221M}, Cen\,A-MM-dw1/Cen\,A-MM-dw2 \citep{2014ApJ...795L..35C}, dw1252-40/dw1251-40 \citep{2017A&A...597A...7M}, and dw1243-42b/dw1243-42 \citep{2017A&A...597A...7M}. The two pairs at the bottom  don't have distance estimates and are therefore uncertain satellite-of-satellite systems.}
\label{tab:observed}
\end{table}

In Fig.\,\ref{fig:ratios} we plot the luminosity ratio of the satellite-of-satellite systems as a function of the brighter dwarf stellar mass. Furthermore, we add six satellite-of-satellite systems from the Feedback in Realistic Environment (FIRE)/GIZMO hydrodynamic zoom-in simulations of isolated dark matter halos \citep{2015MNRAS.453.1305W}, which span host stellar masses between 10$^4$ to 10$^6$\,M$_\odot$. Half of these systems have a similar stellar mass as CenA-MM-dw1 {-- the faintest main dwarf in our sample -- } and half are fainter. 
\citet{2015MNRAS.453.1305W} give the stellar masses of the brightest satellites of the simulated dwarfs. Generally, they find higher luminosity ratios ($\approx$1000) than what is observed in the local universe ($<$100). However, one system at a lower stellar mass than observed has a luminosity ratio of three. {Also using FIRE, \citet{2019MNRAS.489.5348J} studied the luminosity function of isolated LMC-like galaxies. Based on their Fig.\,4, we estimate that the luminosity ratio between their LMC-like galaxies and the most massive satellite ranges between 1 and 160, which is of the same order of magnitude than for the observed LMC and the other observed dwarfs here (which are at lower stellar masses though).}

\begin{table}[ht]
\caption{The satellite-of-satellite systems from the cosmological zoom-in simulations of \citet{2018A&A...616A..96R}.}             
\centering                          
\begin{tabular}{l l l l l}        
\hline\hline                 
 Model & L$_{*}$  &  L$_{*, 1. satellite}$ & sep. & ratio\\    
 &  10$^6$ L$_{\odot}$ & L$_{\odot}$ & kpc & \\    
\hline      \\[-2mm]                  
h025 &         309.37 &    36467 &   18 &  8484 \\
h019 &         291.39 &    24874 &   15 & 11715 \\
h021 &         233.79 &   270145 &   13 &   865 \\
h076 &          18.56 &     1917 &   19 &  9682 \\
h048 &           8.02 &     2029 &   14 &  3953 \\
h050 &           4.16 &      717 &   12 &  5802 \\
h064 &           0.41 &     1982 &   13 &   207 \\
h039 &           0.19 &    19713 &    9 &    10 \\
h141 &           0.22 &      688 &    5 &   320 \\
h061 &           0.21 &      713 &   15 &   295 \\
h111 &           0.20 &      741 &   14 &   270 \\
h091 &           0.17 &     9007 &   12 &    19 \\
h123 &           0.13 &   161141 &   14 &     1 \\
h180 &           0.12 &     1336 &    5 &    90 \\
\hline
\end{tabular}
\label{tab:simulated}
\end{table}

Another 14 satellite-of-satellite systems are drawn from the cosmological zoom-in simulations of \citet{2018A&A...616A..96R}. 
{We determined those satellite-of-satellites by looking 
for sub-halos hosting stars in the virial halo of the four dwarf models having a luminosity (at z=0) 
between $10^5$ and $10^8\,\rm{L_{\odot}}$. For all of these dwarf halos, we found ultra-faint galaxy satellites associated with them. For the  three hosts resembling the M96-DF6 system in luminosity, namely \texttt{h076}, \texttt{h048}, \texttt{h050} \citep[see Tab.~1 of][]{2018A&A...616A..96R} we find satellites with luminosities of  $2\times 10^3$, $2\times 10^3\,\rm{L_{\odot}}$, $1.8\times 10^3$ and distances to the host dwarf of $27$, $20$ and $17\,\rm{kpc}$. For those models, the luminosity ratio is even larger with values above 1'000. 
There are some dwarf models with  higher masses ($\approx$10$^8$\,M$_\odot$) which have a luminosity ratio of the order of 10'000. For the fainter models, i.e. having around 10$^5$\,M$_\odot$, we find the luminosity ratios are matching better the observations, ranging from 1 to 300{, but are of lower stellar mass than observed}.
It is worth mentioning that owing to the resolution, finding satellite-of-satellite at smaller distance is difficult (r12=0.6kpc).
With only a few systems, especially at the observed stellar mass of the main dwarf, it still remain to be determined whether there is a trend or not, but it is conceivable  that $\Lambda$CDM doesn't produce as bright satellite-of-satellites as observed {for dwarf galaxies between 10$^5$\,M$_\odot$ and 10$^8$\,M$_\odot$}. 

}

We furthermore show the resolution limit of IllustrisTNG50 \citep{2019MNRAS.490.3196P,2019MNRAS.490.3234N}, which has a particle gas mass of 8.5$\times$10$^4$\,M$_\odot$. Assuming we need 10 particles to identify an object as a satellite, we can calculate the minimal luminosity ratio as a function of host mass. In Fig.\,\ref{fig:ratios} systems below this calculated line are in principle resolved in IllustrisTNG50. However, we note that the observed dwarf galaxies are sitting at the edge or above this resolution limit, meaning that they are not resolved in IllustrisTNG50. For our observed galaxies, the simulations would need a stellar mass resolution of two orders of magnitude larger than what is currently available with IllustrisTNG50. 

\begin{figure}[ht]
    \centering
    \includegraphics[width=\linewidth]{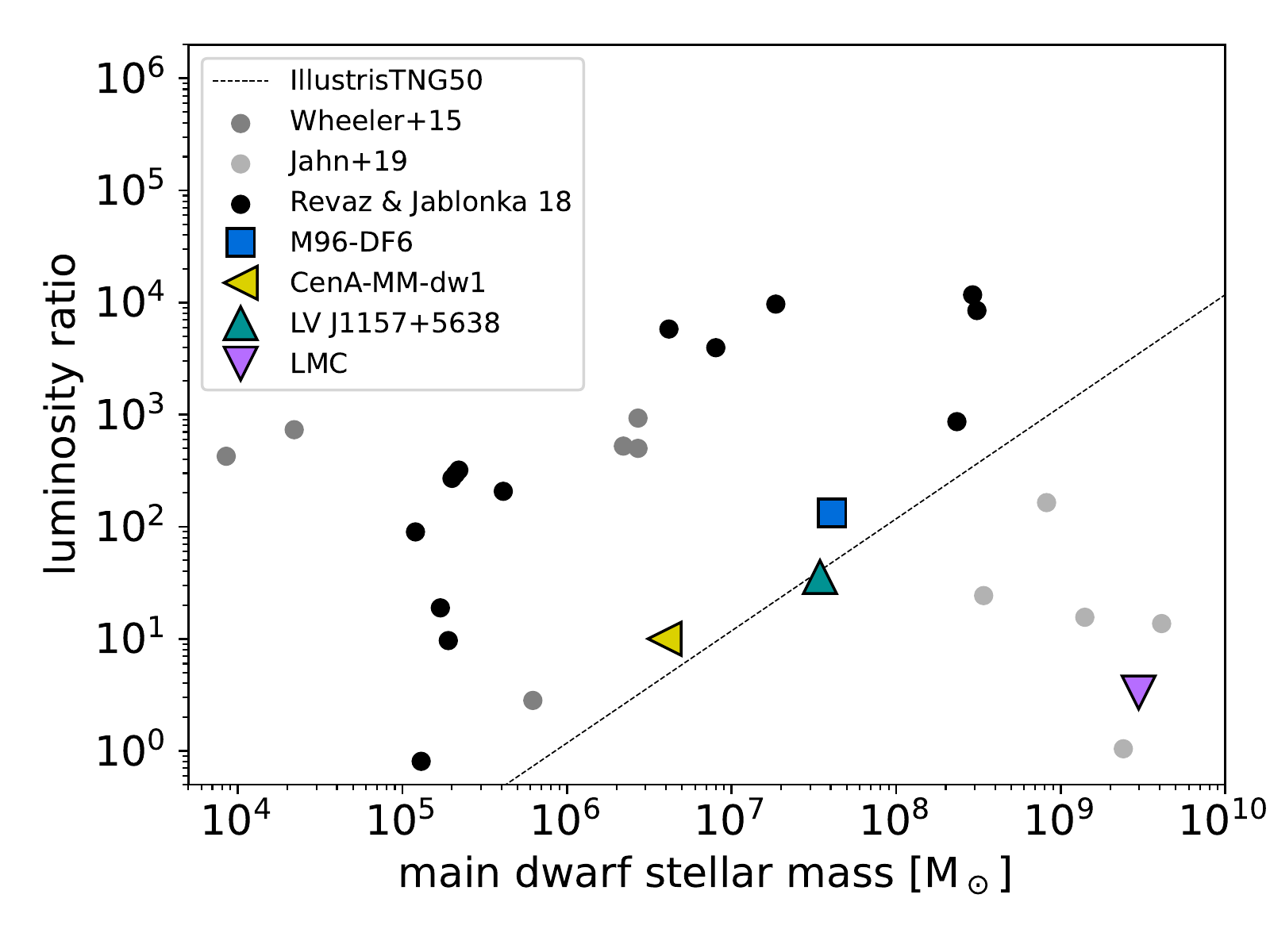}
    \caption{Simulated and observed luminosity ratios of satellite-of-satellite systems as a function of the luminosity of the main dwarf. The gray dots correspond to the {six} satellite-of-satellite systems {analysed} in the FIRE simulation \citep{2015MNRAS.453.1305W}, the black dots to the 14 satellite-of-satellite systems identified in the zoom-in simulations of \citet{2018A&A...616A..96R}. The colored square and triangles {are measurements from} observed satellite-of-satellite systems. The line indicates the resolution limit of the Illustris-TNG50 simulation.}
    \label{fig:ratios}
\end{figure}

\section{Summary and conclusions}
Going through the Extragalactic Distance Database {we established a sample of 117 dwarf galaxies with available HST observations and with distances 8\,Mpc or larger}. We searched for dwarf galaxy satellites of dwarf galaxies and in these HST images found one known satellite-of-satellite for LV J1157+5638 and one previously unknown dwarf galaxy in the {immediate vicinity}  of M96-DF6 -- an ultra-diffuse galaxy in the nearby Leo-I group -- as an extremely faint overdensity of stars. Using a growth curve to measure the structural parameters of this new dwarf -- dw1046+1244 -- we find it to be an ultra-faint dwarf with a stellar luminosity of 1.5 $\times$ 10$^5$\,L$_\odot$ and an effective radius of 205\,pc. this is similar to other ultra-faint dwarfs in the Local Group such as AndXXX. The newly discovered dwarf follows the size-luminosity relation as defined by the Local Group dwarf and is 100 times fainter than its host. Other known satellite-of-satellite systems in the nearby universe share similar properties in terms of their separation (2-20\,kpc) and luminosity ratio (2-135). Noteworthy is that two out of the four satellite-of-satellite systems have an ultra-diffuse galaxy as central dwarf. It has been suggested that some ultra-diffuse galaxies are failed galaxies \citep{2016ApJ...828L...6V} and therefore host a vast dark matter halo. If this is the case, it could explain why they have a -- still relative massive - satellite associated to them.

There exist only a few cosmological simulations which can resolve dwarf galaxies and their dwarf galaxy satellites {to date}. One of them is FIRE, a  hydrodynamic zoom-in simulations of isolated dark matter halos.  \citet{2015MNRAS.453.1305W} simulated six such satellite-of-satellite systems, of which five have luminosity ratios of the order of 1000, {a factor 10} larger than what is observed in the nearby universe.
{Again using FIRE, \citet{2019MNRAS.489.5348J} looked at five isolated LMC-like systems and their satellites. Extracting the luminosity of their brightest satellites, we found that these systems have luminosity ratios between 1 and 160, which is in the range of the observed dwarf systems here. However, apart from the LMC, these simulated systems are all much more massive than the observed dwarfs.}

Another set are the cosmological zoom-in simulations of \citet{2018A&A...616A..96R}. Within 14 of their dwarf galaxy models, we searched for their most luminous satellite. For the more massive dwarfs ($>$10$^6$M$_\odot$), we find luminosity ratios much larger than 1'000. For the less massive dwarf models, the ratios range between 1 to 300, which is closer to our observations. However, these latter dwarf models are all less luminous than the observed dwarfs we consider here.
Only three of the dwarf models have a similar stellar mass and they have all much larger luminosity ratios than observed, which is consistent with the simulated dwarfs from \citet{2015MNRAS.453.1305W} for this mass range. {It is interesting to note that the three brightest hosts from \citet{2018A&A...616A..96R} have similar stellar masses as the faintest hosts from the FIRE simulation in \citet{2019MNRAS.489.5348J}, but rather different luminosity ratios. This could indicate that different simulations produce different satellite systems. However, this would then open the question whether $\Lambda$CDM has solved the abundance problem, when the predicted luminosity functions are depending on the simulations.}  

{For three out of four observed dwarf galaxy satellite systems, cosmological simulations do not find produce the observed luminosity ratio between the dwarfs and their brightest satellites. If the two candidate satellite systems around Cen\,A are confirmed to be true satellites, they will further strengthen this picture.} 
It is {therefore} {conceivable} that in $\Lambda$CDM the brightest satellites of dwarf galaxies are missing, {at least in the regime of the classical dwarfs}, reminiscent of the too-big-to-fail problem (in reverse). 
The problem here is not that we do not observe dwarf satellite systems with a large luminosity ratio, but that we do observe satellite-of-satellite systems with a low luminosity ratio, which we do not find in cosmological simulations (at given stellar mass). More observations and more detailed cosmological simulations of dwarfs at the observed stellar mass are needed to assess whether there is a satellite-of-satellite (SoS) problem, but the data and two independent sets of simulations indicate that this may be the case.

\label{summary}

\begin{acknowledgements} 
{We thank the referee  Guillaume Thomas for the constructive report, which helped to clarify and improve the manuscript.}
O.M. and N.H. are grateful to the Swiss National Science Foundation for financial support under the grant number  	PZ00P2\_202104. {O.M. likes to thank Mridul K Thomas (@mridulkthomas)  for pointing out on Twitter the online tool automeris (https://apps.automeris.io/wpd/) to extract data points from figures, which we used to reproduce the values of \citet{2019MNRAS.489.5348J} in Fig.\,\ref{fig:ratios}.}
\end{acknowledgements}

\bibliographystyle{aa}
\bibliography{bibliographie}

\end{document}